\documentclass{article}
\usepackage[T1]{fontenc} 
\usepackage[utf8]{inputenc} 
\usepackage{ismir,amsmath,cite,url}
\usepackage{graphicx}
\usepackage{color}
\usepackage{url}

\usepackage[most]{tcolorbox}

\newtcolorbox{greybox}{
  colback=gray!10,
  colframe=gray!50,
  boxrule=0.4pt,
  arc=1mm,
  left=7pt,
  right=7pt,
  top=7pt,
  bottom=7pt
}

\renewcommand{\permission}{}

\usepackage{lineno}

\title{Harmonic and Transposition Constraints Arising from the Use of the Roland TR-808 Bass Drum}

\oneauthor
 {Emmanuel Deruty}
 {Sony Computer Science Laboratories, Paris, France; Aalborg University, Denmark \\ {\tt emmanuel.deruty@sony.com}}

\def\authorname{E. Deruty}

\begin{document}

\maketitle
\begin{greybox}
\begin{small}
Published version: Deruty, E. (2024).
Harmonic and Transposition Constraints Arising from the Use of the Roland TR-808 Bass Drum.
\textit{Proceedings of the 25th International Society for Music Information Retrieval Conference}, 78--85.
San Francisco, California, USA and Online. \\\url{https://doi.org/10.5281/zenodo.14877286}
\end{small}
\end{greybox}
%

\vspace{1cm}

\begin{abstract}



The study investigates hip-hop music producer Scott Storch's approach to tonality, where the song's key is transposed to fit the Roland TR-808 bass drum instead of tuning the drums to the song’s key. This process, involving the adjustment of all tracks except the bass drum, suggests significant production motives. The primary constraint stems from the limited usable pitch range of the TR-808 bass drum if its characteristic sound is to be preserved. The research examines drum tuning practices, the role of the Roland TR-808 in music, and the sub-bass qualities of its bass drum. Analysis of TR-808 samples reveals their characteristics and their integration into modern genres like trap and hip-hop. The study also considers the impact of loudspeaker frequency response and human ear sensitivity on bass drum perception. The findings suggest that Storch’s method prioritizes the spectral properties of the bass drum over traditional pitch values to enhance the bass response. The need to maintain the unique sound of the TR-808 bass drum underscores the importance of spectral formants and register in contemporary popular music production.


\end{abstract}


\section{Introduction}
\label{sec:intro}



In popular music, a common practice is to tune the drums to the song's key \cite{toulson2009perception}. However, in a 2007 interview \cite{zisook2007}, R\&B producer Scott Storch suggests that during the production of music involving a Roland TR-808 drum machine, it may be beneficial to do the opposite and transpose the song's key to fit the 808 bass drum \cite{storch2007}. The process involves transposing all the tracks but the bass drum. Storch's motive for undertaking such a potentially time-consuming set of operations is to conserve the characteristic sound of the 808 bass drum.

 
 The present study investigates aspects of the music production process that may explain Storch's position. In Section~\ref{sec:drumsandtuning}, we shortly address the issue of drum tuning in popular music. In Section~\ref{subsec:808andproduction}, we provide an overview of the importance and usage of the Roland TR-808 in popular music production, focusing on its bass drum voice. In Section~\ref{subsec:808analysis}, we analyze the content of TR-808 bass drum samples. In Section \ref{subsec:subbassand808}, we relate spectral features of TR-808 bass drum samples to a diachronic analysis of the power spectrum in popular music. In Section~\ref{sec:tuningto}, we can understand Storch's position by involving the frequency response of loudspeakers and the sensitivity of the human ear. Finally, in Section~\ref{sec:pitchandregister}, we discuss how the practice suggested by Storch may be a particular case of how properties of the spectrum might be considered more important than pitch values.

\section{Drums and tuning}\label{sec:drumsandtuning}

The musical signal has been divided into two categories: ``percussion has a short temporal duration and is rich in noise, while harmonic elements have a long temporal duration with most of the signal energy concentrated in pitch spikes'' \cite{rump2010autoregressive}. ``The harmonic and percussive components of music signals have much different structures in the power spectrogram domain, the former is horizontal, while the latter is vertical'' \cite{ono2008real}. These observations are the basis for source separation methods distinguishing “drums” from “pitched instruments” \cite{fitzgerald2010harmonic,yoo2010nonnegative}.

Yet, drums can contain pitched content \cite{richardson2010acoustic}. In drum sounds, relations between eigenfrequencies are not necessarily harmonic \cite{antunes2017possible}. “The tonal elements in drums are usually not structured like partials in a harmonic series. Instead, their frequency relationship can range from inharmonic to chaotic” \cite{wu2018review}. From a music producer's perspective, ``drums make several different notes simultaneously'' \cite{Roberts}.

Recent source separation methods do not involve prior hypotheses. They're based on models trained on actual data. Listening to the audio output stemming from such technology indicates that drum stems extracted from popular music do contain pitch. Demonstrations of Steinberg's SpectraLayers \cite{Attack2022}, Native Instruments' iZotope RX 8 \cite{Attack2022}, iZotope RX 9 \cite{Loose2022}, and StemRoller \cite{Stemroller2023}, provide relevant examples. 

\newpage

If drums contain pitched content, they can be tuned. In popular music, the ``[i]ntricate tuning of acoustic drums can have a significant impact on the quality and contextuality of the instrument'' \cite{toulson2009perception}. 
There is no consensus on how to tune drums: ``[t]alk to ten different drummers and you’ll get ten different ways to tune drums [...] there’s actually no wrong or right way to tune a drum, or right or wrong pitches to tune it to'' \cite{drummagazine}. 

Scott Storch is an American record producer and songwriter. Storch has been referred to as a ``producer that changed the R\&B game'' \cite{Singleton2022}, a ``superproducer'' \cite{chapman2008ill}, \textit{i.e.} a wave of artists ``who have established a new degree of visibility for the rap producer, earning star billings virtually equal in prominence to the artists that they produce'' \cite{Levine2003}. For Scott Storch, drum tuning is an integral part of the music production process:

\begin{quote}
``I know there’s a lot of producers [who will] put an 808 in the song, and there will be chords and stuff clashing with it, and [...] if [...] your ears are really in tune with that stuff, you realize it’s just like [``not so convincing'' kind of gesture]... Sometimes, it actually does something cool to the track, but [...] I like to [...] get into that and tune the kick to match [...] the bass line or whatever the chords are doing [...], I just try different stuff... then... [even when there is] not an incredible amount of tune that carries over regular kicks, like short kicks, and I find myself sometimes at least even trying to tune [...] a regular [...] kick drum sound, and get it close to where most of the chords are in the song...'' \cite[0:30]{zisook2007}
\end{quote}

In the above, Storch mentions the Roland TR-808 bass drum and testifies to tuning bass drums to match the music's key.

    \section{The Roland TR-808}\label{subsec:808andproduction}


The Roland TR-808 Rhythm Composer is an analog drum machine manufactured between 1980 and 1983 \cite{hasnain2017tr}. It is ``one of the most influential and unique drum machines of its time'' \cite{meyers2003tr}. ``To this day, the 808 remains a benchmark against which all other analog drum machines are measured'' \cite{werner2014physically}. It can be found in many music genres. The TR-808’s distinctive presets are classic sounds in hip-hop, techno, electro, R\&B, and house music \cite{dayal2014tr}. The 808 ``play[ed] a central role in the development of acid house''\cite{werner2014physically}. Pop music star Phil Collins used it throughout his entire career \cite[1:21:28]{Dunn2015}. It is ``a fixture in hip-hop culture, not only as a tool for producers but as a defining sound of the genre'' \cite{hasnain2017tr}. According to Scott Storch, in modern trap music, producers ``live in an 808 world'' \cite{storch2022}. One reason for the success of the 808 resides in the fact that ``it sounded like nothing else [...] and this is what made it so distinctive'' \cite{carter1997tr}. 
Perhaps as a result, the 808 has been seen not only as a drum machine but as an ``instrument in its own right'' \cite[0:06:51]{Dunn2015}.


One notable voice of the 808 is its ``long and velvet deep, almost subsonic'' bass drum \cite{carter1997tr}, which can be made into a ``multi-second-long decaying pseudo-sinusoid with a characteristic sighing pitch'' \cite{werner2014physically}. According to producer Pharrell Williams, the 808 bass drum ``filled a massive void in the sound spectrum that was not there [...] once the 808 started to occupy that space, it became like something that you missed if you did not have it'' \cite[1:20:52]{Dunn2015}.

Over time, the 808 bass drum became used as both kick drum and bass. According to producer Remi Kabaka Jr., ``the kick drum would play the bass at the same time [...] there was drums and there was bass, but now the two were sort of fused, so the fill was not just complex and rhythmical, but it was also tonal'' \cite[1:11:11]{Dunn2015}. Musician and writer Alex Lavoie notes that ``[i]n most contemporary music genres, especially in trap and hip-hop, the 808 often carries the bassline, providing both the low-end foundation and outlining the harmonic progression of the song'' \cite{lavoie2020}. Musician and producer Charles Burchell writes that the TR-808 ``brings a sound closer to a traditional bass line while retaining the power of a drum [...]  In many cases, producers will not use a kick drum sample. Instead, they program drum patterns with a tuned 808 as the kick drum'' \cite{burchell2022}.

As a tonal instrument, the 808 bass drum can be tuned: as Lavoie states, ``[a]n 808 kick, particularly when it has a long decay, effectively functions as a bass instrument. That’s why tuning your 808s is so crucial'' \cite{lavoie2020}. 
Lavoie warns that ``[i]f the pitch of your 808 kick doesn’t match the key of your song, it can create a dissonant effect'' \cite{lavoie2020}.



\vspace{.5cm}
\section{The 808 bass drum}\label{sec:808bassdrum}

\subsection{Signal analysis of 808 bass drum samples}\label{subsec:808analysis}

\begin{figure}[htbp]
  \centering
  \includegraphics[width=.9\columnwidth]{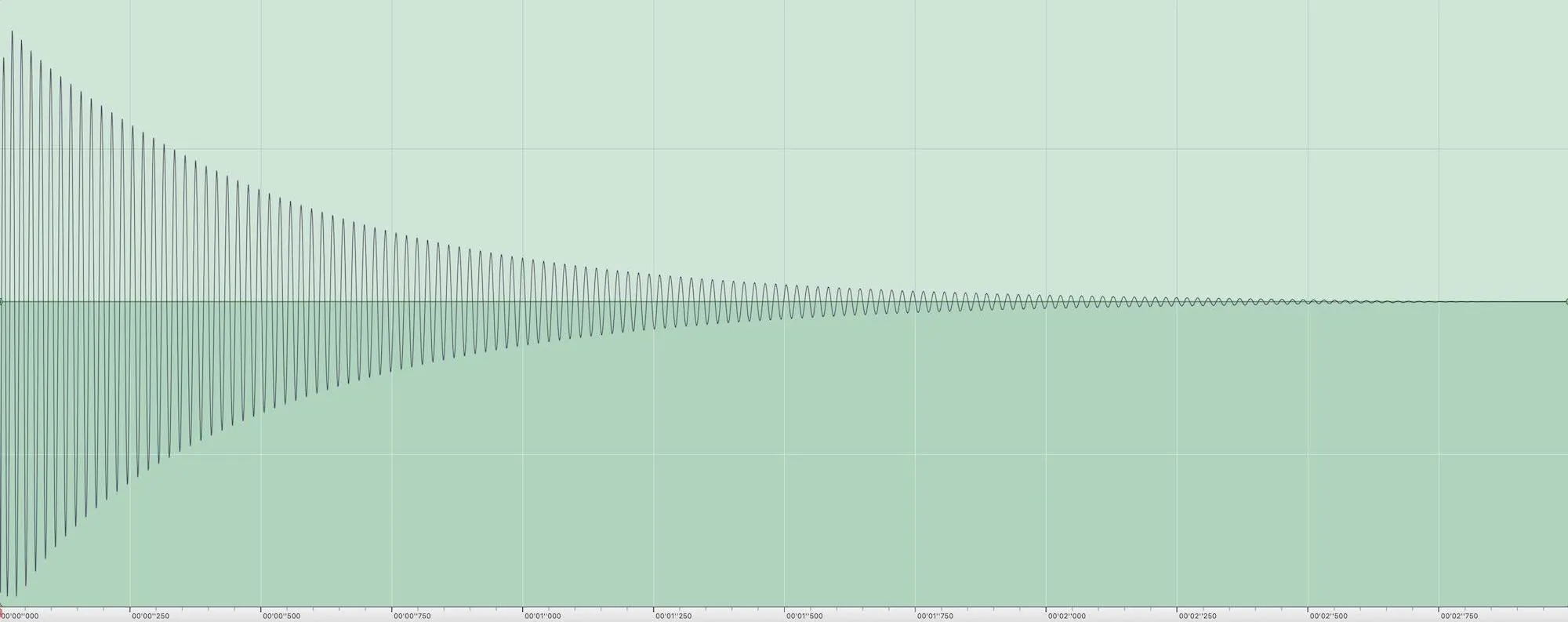}
  \caption{\it ``TR808 BD Bass Drum Long 01'' sample, waveform.}
\label{fig:TR_Waveform}
\end{figure}

Figure~\ref{fig:TR_Waveform} shows the waveform corresponding to the ``TR808 BD Bass Drum Long 01'' preset. All samples considered in this paper originate from the TR-808 Trisample library \cite{trdownload}. The waveform confirms that the sample is tonal. The tonal aspect derives from the TR-808 generation technique, during which an oscillator produces a sawtooth wave that is filtered to make it close to a sine wave \cite{reid2002bass}.

\newpage

\begin{figure}[htbp]
  \centering
  \includegraphics[width=1\columnwidth]{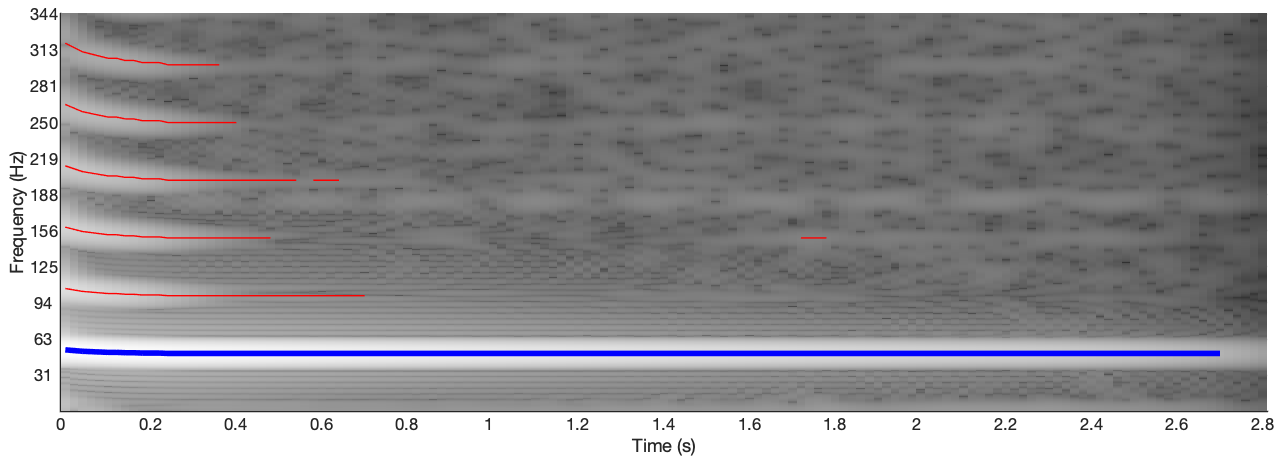}
  \caption{\it``TR808 BD Bass Drum Long 01'' sample, STFT. The horizontal lines follow the fundamental and harmonics. The blue line stops when the energy of the corresponding bin is lower than 0.7 times the peak energy of all bins. The red lines stop when the energy of the corresponding bin is lower than 0.5 times the peak energy of all bins. }
\label{fig:TR_STFT}
\end{figure}

Figure~\ref{fig:TR_STFT} shows the STFT for the same sample. Harmonics are present near the start of the sample and then fade out. The sample’s pitch value is briefly higher near the beginning, then decreases to a stable value. A study of the 37 ``long'' samples from the Trisample library shows that the median range for the initial frequency sweep is close to one half-tone.

\begin{figure}[htbp]
  \centering
  \includegraphics[width=1\columnwidth]{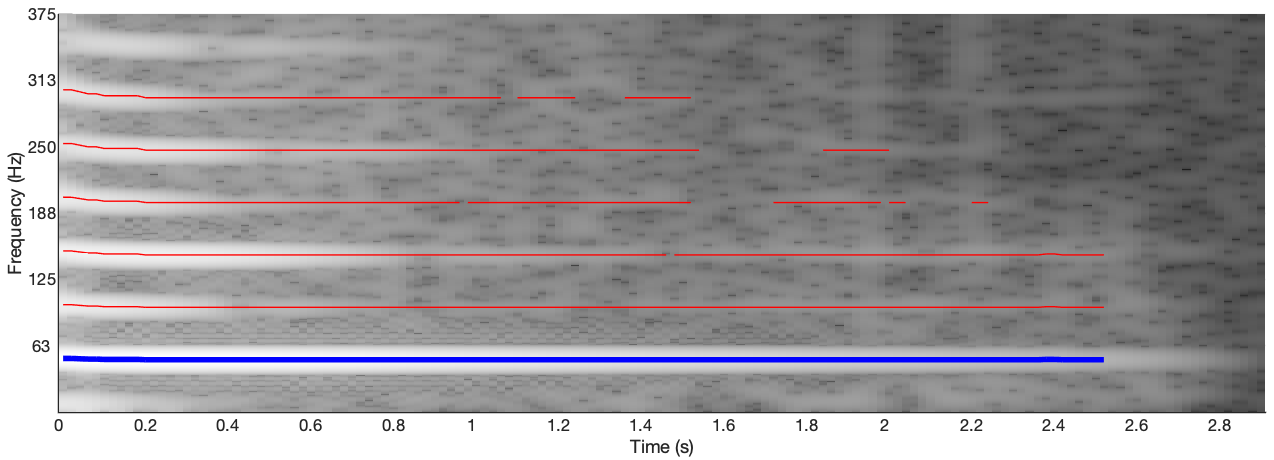}
  \caption{\it ``TR808 BD Bass Drum Driven 01'' sample, STFT.}
\label{fig:TR_driven_STFT}
\end{figure}

 The Tristar library features ``driven'' samples (a reference to the slang term ``drive'' for ``overdrive'', \textit{i.e.} ``distortion''). Figure~\ref{fig:TR_driven_STFT} shows the STFT for one ``driven'' sample. The threshold conditioning the display of the partials as red lines is the same as in Figure~\ref{fig:TR_STFT}, which indicates that the distortion boosts the level of the overtones.

\begin{figure}[htbp]
  \centering
  \includegraphics[width=1\columnwidth]{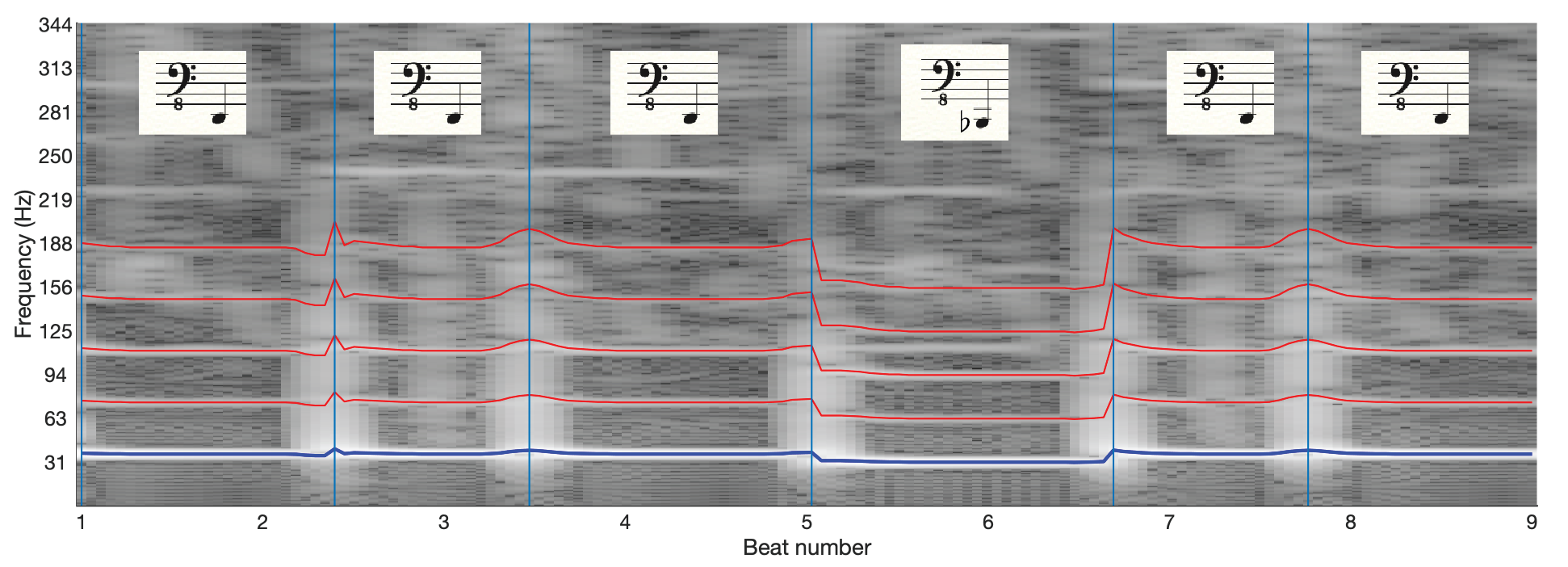}
  \caption{\it Future, ``Mask Off'', 8 beats from 0'25 to 0'30, STFT. The vertical lines denote the kick drum’s onsets. The horizontal lines follow the TR’s fundamental and harmonics. The corresponding pitch values are shown at the top.}
\label{fig:Mask_off}
\end{figure}

Figure~\ref{fig:Mask_off} shows the STFT for an extract from the 2017 song ``Mask Off'', by the American rapper Future. The track has been described as an example of heavy 808 use~\cite{hasnain2017tr}. The initial frequency sweep on each bass drum occurrence is similar to the samples shown in Figures~\ref{fig:TR_STFT} and~\ref{fig:TR_driven_STFT}. The vertical distribution of high energy values at the beginning of each bass drum occurrence suggests that the 808 is superimposed with a noisier kick drum. The 808 samples are tuned to the song’s tonality (D minor). The pitch values (D1 and B$\flat$0) are very low: they stand one minor second and one perfect fourth above the piano's lowest note. The corresponding frequency range (ca. 40Hz) recalls the ``almost subsonic'' aspect of the 808 bass drum samples \cite{carter1997tr}.


    
    
    
    
    

\vspace{.5cm}
\subsection{Sub-bass frequencies and the 808 bass drum}\label{subsec:subbassand808}

\begin{figure}[h] 
  \centering
  \includegraphics[width=1\columnwidth]{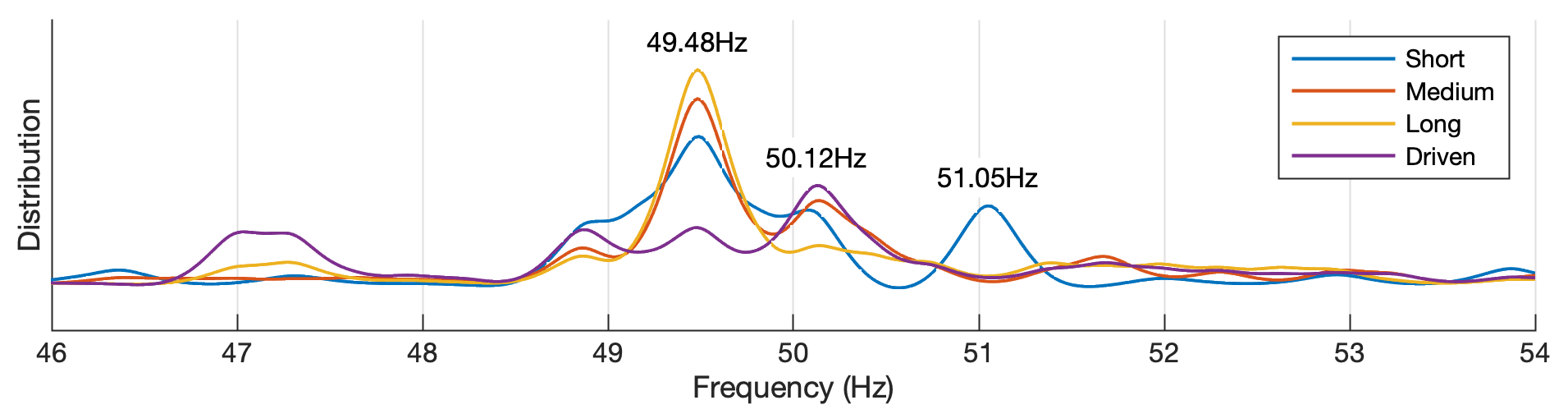}
  \caption{\it Distribution of fundamental frequencies of TR-808 bass drum samples. The fundamental frequencies are evaluated on 0.2-second windows. The contribution of each window is weighted according to the energy at the fundamental frequency. In the non-``driven'' presets, the maximum of the distribution corresponds to $f_0 = 49.48$~Hz. The $f_0$ values for the ``driven'' presets are higher. ``Short'' presets involve a secondary local maximum ($f_0=51.05$~Hz) corresponding to the samples' earliest windows.}
\label{fig:Dists}
\end{figure}

Producers recognize three distinct regions of sub-bass: the ``boom'' (ca. 30Hz), the ``thump'' (ca. 50Hz) and the ``punch'' (ca. 80Hz) \cite[pp. 88--118]{fink2018relentless}\cite[p. 282]{fink2020boom}.  Figure~\ref{fig:Dists} confirms that 50Hz (the ``thump'') is the ``frequency range occupied by the Roland TR-808 analog kick'' \cite{fink2020boom}.

\begin{figure}[h] 
  \centering
  \includegraphics[width=1\columnwidth]{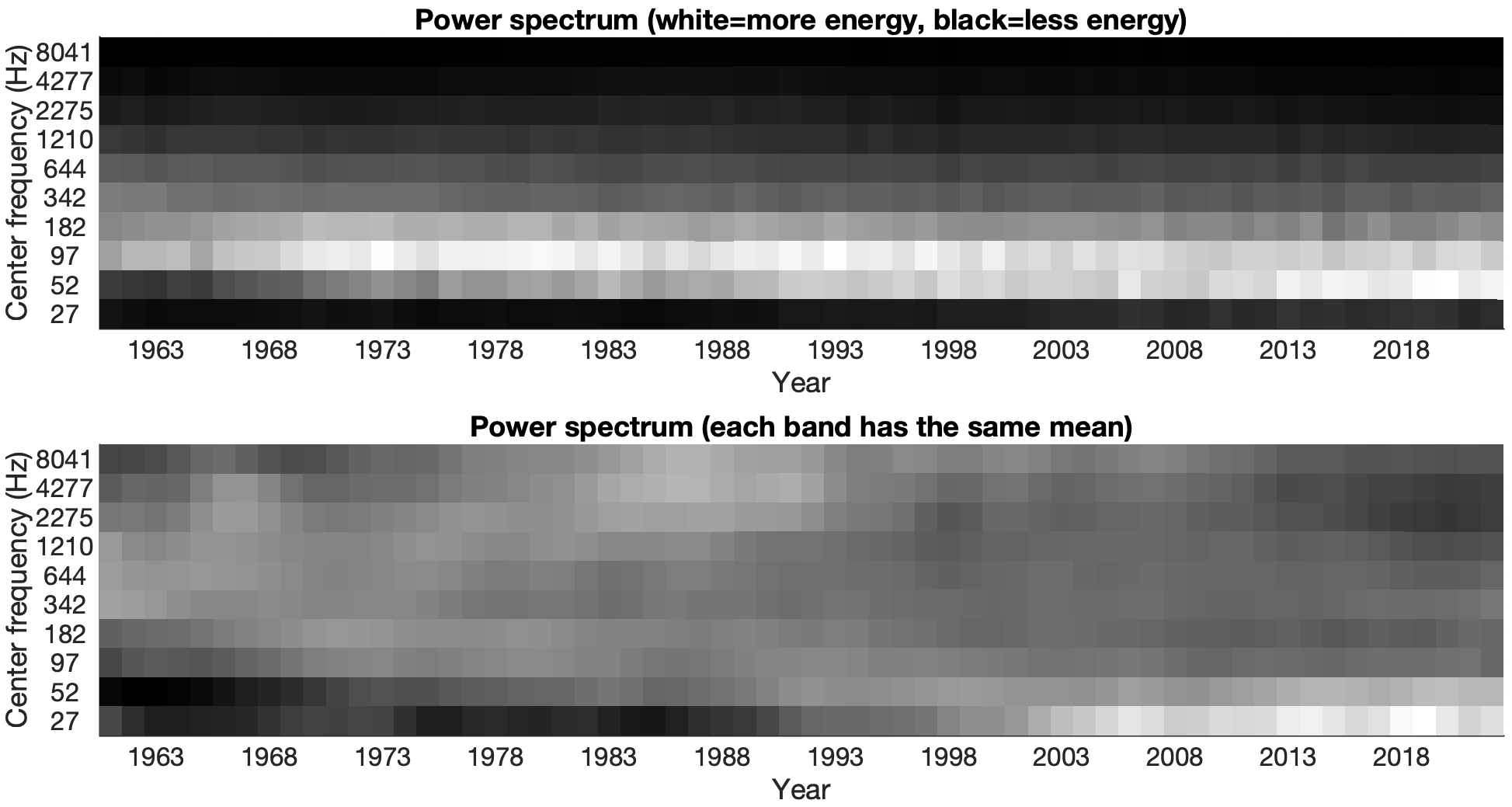}
  \caption{\it Evolution of the power spectrum in popular music. Top, raw energy values. Bottom, values for each frequency band are normalized to the same mean.}
\label{fig:specs}
\end{figure}

Before the advent of digital audio, low frequencies were attenuated to protect amplifiers and speakers from the adverse effects of mechanical noise and harmonic distortion \cite[p. 282]{fink2020boom}\cite{read1952reproduction,millard2005america}. Musical information in this frequency range only became possible by using digital audio as a medium. Figure~\ref{fig:specs} shows the evolution of the power spectrum in popular music. The measures were derived from a dataset containing 30435 tracks released between 1961 and 2022. The choice of the tracks stems from the ``Best Ever Albums'' website, a review aggregator that proposes the best-rated albums for each year of production \cite{BEA}. For each year, we select the best-rated albums. The overall spectral profile is consistent with Pestana's results \cite{pestana2013spectral}. The increase of energy in the lower band, also testified by Hove et al. \cite{hove2019increased}, is concomitant to the advent of digital audio. 



\newpage

The analysis results shown in Figure~\ref{fig:TR_STFT} indicate that after the initial 0.4s-long attack, ``long'' 808 samples are based on a single low-frequency sine wave. The sine wave's frequency is ca. 50~Hz according to  Figure~\ref{fig:Dists}. The results shown in Figures~\ref{fig:TR_driven_STFT} and~\ref{fig:Mask_off} suggest that this very low frequency remains an essential component of the 808 bass drum with added harmonics. Confronting these observations with the power spectrum evolution in popular music (Figure~\ref{fig:specs}), it follows that the sound of the 808 bass drum was not fully reproduced before the end of the '90s, even though the machine itself was sold between 1980 and 1983.


After 2009, ``the characteristic 808-kick drums [...] started entering mainstream music in general'', and trap music, a ``tradition of rap that developed during the 1990s'', an ``808 world'' according to Scott Storch (see Section~\ref{subsec:808andproduction}), ``began to reach strong presence on the mainstream Billboard music charts'' \cite{kaluvza2018reality}. So strong is the presence of trap in the charts that this formerly underground genre has been qualified as ``pop'', in the sense that ``[p]eople’s ears have adjusted'' to it \cite{Lee2017}.

The extended bandwidth provided by the emergence of digital audio made possible the faithful restitution of the entire spectrum of the 808 bass drum, which favored the birth and rise of a music genre that became mainstream and influenced popular music in general. 

\section{Tuning the song's key to the TR-808 bass drum}\label{sec:tuningto}

In Section~\ref{sec:drumsandtuning}, Scott Storch describes how he tries to tune the bass drum (808 in particular) to the music's key. Later in the same interview, Storch suggests that instead of tuning the 808 bass drum sample to the song's tonality, one can do the opposite and adjust the song's key to the 808 bass drum sample:

\begin{quote}
``[S]ometimes, producers will program a song in a certain key, and they’ll try to program an 808 under it, and it’s like the key of the song is almost too low to really let speakers do what they need to do with the bass so, I recommend [...] modulating the song up, transposing it up a couple of keys, and you’ll be surprised how much more level you can get out of the song. [Because] anything really below […] a low E [...], it’s like the speakers are gonna not, let you turn it up, you don’t feel the bass response.'' \cite[1:37]{zisook2007}
\end{quote}

Storch describes a situation in which a producer previously set the key for a song, tunes an 808 bass drum to make it fit the key, and, as a result, the 808 bass drum does not sound ``right''. 

\vspace{.5cm}
\subsection{Transposition of the TR-808 bass drum: effect on the lowest partial}\label{subsec:fundamental}

Let us consider an example where the song's key is D, as in the extract from Figure~\ref{fig:Mask_off}. We focus on the fundamental, the only lasting component in samples from the ``long'' type (Figure~\ref{fig:TR_STFT}). As seen in Figure~\ref{fig:Dists}, the $f_0$ of an 808 bass drum is ca. 49.5Hz, corresponding to a G1. The producer, therefore, transposes the 808 bass drum one perfect fourth down (5 semitones) to a D1 -- one tone below the ``low E'' mentioned by Storch. Storch states that the loudspeakers may not reproduce the bass correctly in such a situation.


Professional mixing engineers mainly use near-field monitors \cite[p. 3]{senior2011mixing}. With such monitors, they can produce ``masters which `travel' well to their use by the record buyers'' \cite{newell2001yamaha}. The use of near-field monitors extends to producers. 
Nigel Godrich testifies that during the production of Radiohead's ``OK Computer'', he always used near-field monitors, but never the studios' main monitors, which ``don't relate to anything'' and are ``fairly useless'' \cite{Robinson1997}. In the many videos documenting his work, Storch can be seen using near-field monitors. 

\begin{figure}[htbp]
  \centering
  \includegraphics[width=1\columnwidth]{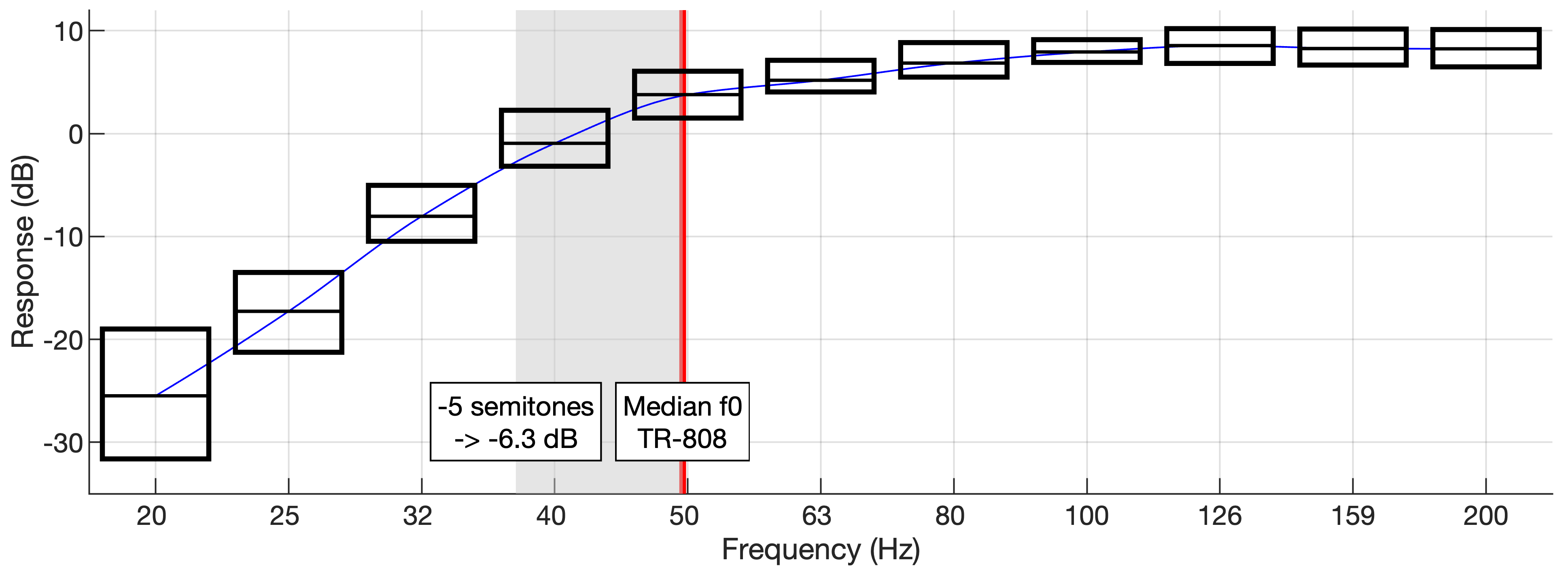}
  \caption{\it near field loudspeaker responses as a function of frequency. The center horizontal line in each box represents the median, and the two surrounding horizontal lines represent the 25\textsuperscript{th} and 75\textsuperscript{th} percentiles. The blue line shows the smoothed median response. The red vertical line represents the median $f_0$ for the 808 bass drum as shown in Section~\ref{subsec:subbassand808}, Figure~\ref{fig:Dists}. The gray rectangle denotes a -5 semitone transposition of the median $f_0$. The textual representation displays the difference in the response that occurs.}
\label{fig:Responses}
\end{figure}

Newell et al. \cite{newell2001yamaha} provide the frequency response for 36 near-field monitoring loudspeakers. Figure~\ref{fig:Responses} graphs the median frequency response for these loudspeakers against the median $f_0$ for the 808 bass drum samples (49.5Hz / G1) and the TR bass drum median frequency transposed down one perfect fourth (37Hz / D1). The downward transposition results in a gain loss of 6.3 dB. Following Storch's suggestion and transposing up the song key instead of transposing down the 808 sample would avoid the 6.3dB loss. In Storch's terms, transposing the song up may ``let speakers do what they need to do with the bass''.

Loudspeakers are not the only frequency-dependent transducers involved in the listening process. The human ear is also sensitive to frequency. In particular, as the frequencies get closer to the lower limit of human hearing, a sine wave with the same sound pressure level but a lower frequency will be perceived as less loud.

\newpage

\begin{figure}[htbp]
  \centering
  \includegraphics[width=1\columnwidth]{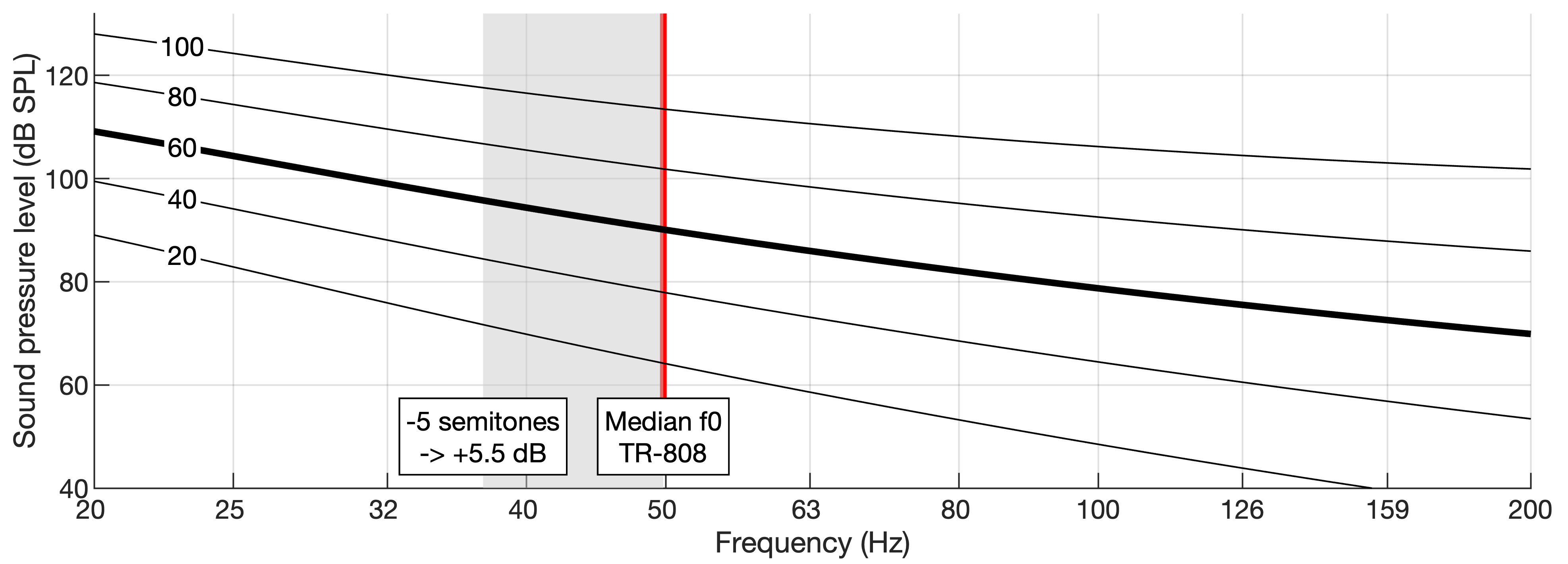}
  \caption{\it Equal-loudness contours according to \cite{iso2262003}. The numbers superimposed on each contour indicate the loudness value corresponding to the contour (in phon). The red vertical line represents the median $f_0$ for the 808 bass drum as shown in Section~\ref{subsec:subbassand808}, Figure~\ref{fig:Dists}. The gray rectangle indicates a -5 semitone transposition of the median $f_0$ (one perfect fourth down). The textual representation displays the gain that would be required so that the transposed $f_0$ remains at the same 60-phon loudness.}
\label{fig:ISO}
\end{figure}

The phenomenon is described by equal loudness contours, representing the sound pressure levels at different frequencies that are perceived as equally loud \cite{fletcher1933loudness}. Figure~\ref{fig:ISO} graphs the ISO226-2003 \cite{iso2262003} equal loudness contours against the median 808 bass drum $f_0$, and the same frequency transposed down one perfect fourth. If we choose a loudness of 60 phon, a +5.5dB gain would be required so that the transposed $f_0$ remains at the same loudness. Therefore, considering the human ear as one of the transducers in the signal path, the gain it applies to the signal when transposing down the original median $f_0$ is ca. -5.5dB. As a result, the overall gain loss following the downward transposition originating from both the loudspeakers and the ear can be estimated to be ca. 11.8 dB. 

\vspace{.5cm}

\newpage

If different 808 bass drum notes result in different gains, then a sequence of different 808 bass drum notes will result in gain changes within the sequence. Quoting Storch, ``for 808s […] I try to stay in the comfort zone of the speaker, so I don’t [...] have the volumes jumping out for different notes'' \cite{Storch2022b}. In other words, 808 bass drum parts' pitch should remain largely static to achieve a stable gain. In turn, largely static bass pitch values may result in a limited variety of chords. The phenomenon illustrates how loudness stability may take precedence over harmonic complexity.


\vspace{.5cm}
\subsection{Transposition of the TR-808 bass drum: involvement of the harmonics}

\begin{figure}[htbp]
  \centering
  \includegraphics[width=1\columnwidth]{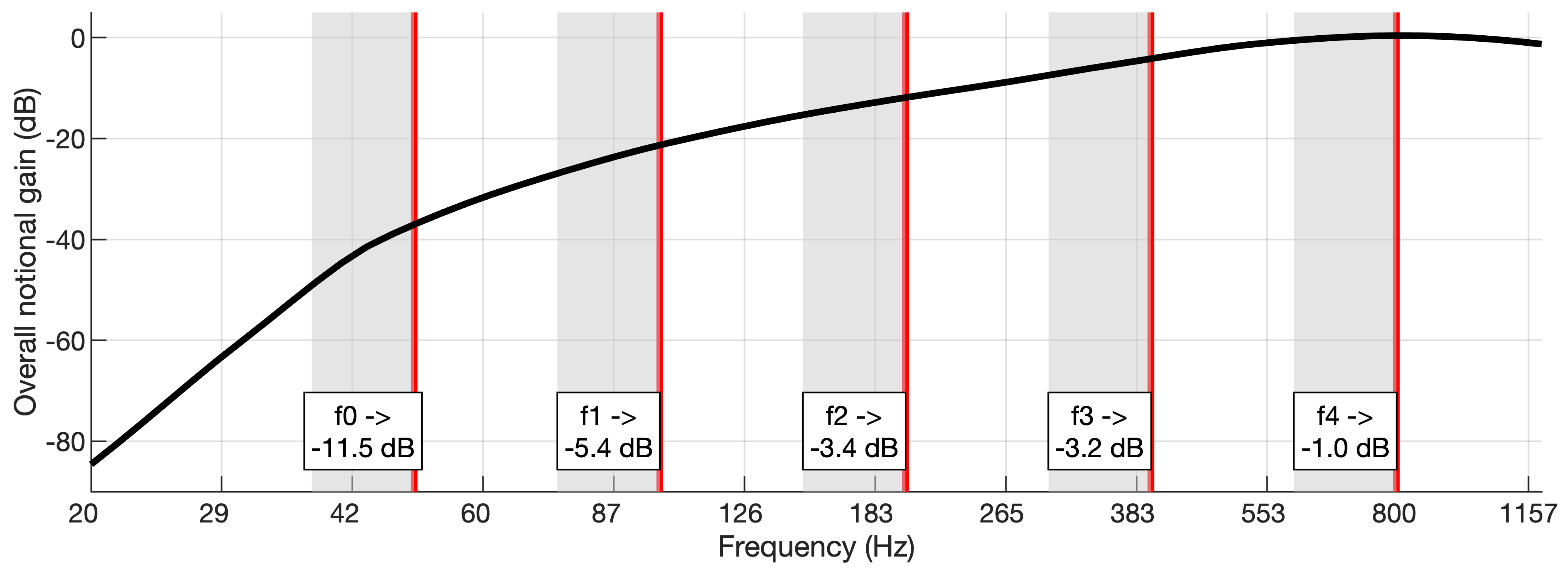}
  \caption{\it The black line shows the combined response deriving from both the near field loudspeakers and the ear's sensitivity at 60 phon (Figures \ref{fig:Responses} and \ref{fig:ISO}). The red vertical lines represent the median values for the 808 bass drum's fundamental and harmonics. The textual representations display the difference in the response that occurs from a -5 semitone transposition.}
\label{fig:both}
\end{figure}

The 808 bass drum samples corresponding to Figures~\ref{fig:TR_driven_STFT} and~\ref{fig:Mask_off} involve lasting harmonics. Figure~\ref{fig:both} shows the combined response deriving from both the near field loudspeakers and the ear's sensitivity at 60 phon. The lower the frequency, the greater the influence of transposition on the overall gain. The gain loss diminishes with each harmonic. It is almost zero for the fifth harmonic.


We generate a 49.5Hz five-partial harmonic complex tone. The amplitudes of the partials are the same as in the ``TR808 BD Bass Drum Driven 01'' sample when the frequency values reach a static regime (see Figure~\ref{fig:TR_driven_STFT}). The overall power change following a 5-semitone downward transposition is -4.5dB. It is much less than the -11.8 dB gain brought by the downward transposition of the lone fundamental. The result suggests that the issues mentioned by Scott Storch (gain conservation and gain stability) mainly concern the fundamental or, at least, the lowest harmonics. In other words, Storch is specifically concerned with the audibility and stability 
of the 808 bass drum's bottom partials. 


\vspace{.5cm}

\newpage



The phenomenon known as the "missing fundamental" \cite{licklider1951duplex} suggests that even if a negative gain is applied to lower harmonics, the perceived pitch remains unchanged due to the auditory system's temporal pitch processing. Only timbre is affected. In the case of the sub-bass register, \textit{i.e.} frequencies lower than 100Hz according to Fink \cite[p. 281]{fink2020boom}, another perceptual aspect may be mentioned. In relation to findings by Takahashi et al. \cite{takahashi2002relationship}, Fink et al. \cite[pp. 88-118]{fink2018relentless} suggest that one aspect of the perceptual effects of bass stems from small body surface displacements. According to the author, each sub-bass range can be associated with a body region in which the corresponding frequencies are imaginatively felt. The ``boom'' (ca. 30Hz) is ``the semi-audible vibration in the gut felt during the deepest drops in dancehall and dubstep''. The ``thump'' (ca. 50Hz) is felt in the stomach, and the ``punch'' (ca. 80Hz) in the chest. Even when listeners use headphones, bass frequencies may be associated with a ``tactile sensation'' \cite{hove2020feel}. Fink \cite{fink2020boom} and Hove et al. \cite{hove2020feel}'s views suggest that low frequencies may play a role beyond pitch and timbre, in this case, a haptic role. 

A downward transposition and the resulting negative gain applied to these frequencies may affect both the resulting timbre and bodily sensation. As a result, they may be prejudicial to at least some music genres, independently from the presence of upper harmonics.

\vspace{.5cm}
\section{Pitch and register}\label{sec:pitchandregister}

Scott Storch's advice according to which a song's key may be adjusted to the 808 bass drum sample is based on the following premise: the transposition of the elements of the music that are not the 808 bass drum is less problematic than the transposition of the 808 bass drum. 
The change in pitch values does not affect the musical intervals, but the shift in register affects the perceived spectral profile. The change in perceived spectral profile is more important in the case of the TR-808 bass drum due to its low-frequency content. 




Following Frisius \cite[p. 81]{frisius2010search}, ``a [music] theory [that] posits a principle of neutral transposition, according to which groups of pitches essentially do not change their character if one transposes them'', does not take into account the transposition of the sounds themselves. Frisius remarks that such a theory may not be suited to music from the 20\textsuperscript{th} century. He mentions the composer Luigi Russolo, who found it difficult to ``transpos[e] sonic gestures into other registers without losing their identity''. According to Frisius, this difficulty is ``felt above all when pitch is not clearly definable''. One way to understand the phenomenon is that transposed melodies are only the ``same as'' each other because they are constructed using a set of pitches whose chromas repeat at the octave. The listener encodes them in terms of pitch sequences. If two sounds are ``transpositions'' of each other but are not perceived in terms of pitch, then they are just different sounds. 


\newpage
Pitch intervals may be robust to transposition, but the register will change, and so will timbre. The phenomenon has been described in orchestration treatises \cite{berlioz1844grand,gevaert1885nouveau,koechlin1954traite}. 
According to Hector Berlioz, as far as the violin is concerned, C major may be ``\textit{grave, mais sourd et terne}'' (rich in low frequencies, but dull and muted), and F\# minor ``\textit{tragique, sonore, incisif}'' (tragic, resonant, incisive) \cite{berlioz1844grand}.
More recently, Reymore et al. \cite{reymore2023timbre} have studied the relation between pitch height and timbre in acoustic instruments. 

Personal interviews with music producers from the production company Hyper-Music (\url{https://www.hyper-music.com/}) suggest that in recent popular music, the simultaneous consideration for pitch values and register when considering transposition is paramount. According to one of Hyper-Music producers, Storch prioritizing the register of an instrument over particular pitch values is a ``basic rule'' of modern music production. Pitch is subservient to spectral formants. Priority is given to the absolute position of the formants in the spectrum. If pitch has to be changed so that the formants of the sound carrying the pitch reach the desired positions, it will be changed. Another producer from the same company claims to be always cautious with transposition, as it may affect timbre. In accordance with Frisius' point of view, if the pitch content of the part is not too strong, the same producer may simply forego transposition, even if the pitched content of the sample conflicts with other tonal elements. 

In productions involving a TR-808 bass drum, Hyper-Music's producers often set the tonality to D or Eb to take full advantage of the bass drum's character. Such tonalities neighbor that of the example shown in Figure~\ref{fig:Mask_off}. 


\vspace{.5cm}
\section{Conclusion}

According to Section~\ref{sec:drumsandtuning}, some authors have previously divided the musical signal into two categories: percussion (rich in noise, short duration), and harmonic elements (long duration, most of the energy concentrated in spikes in the spectrum) \cite{ono2008real,rump2010autoregressive,yoo2010nonnegative}. However, other authors have studied the existence of pitch in percussion \cite{richardson2010acoustic,antunes2017possible,wu2018review}. In music production, drum tuning has been seen as essential \cite{toulson2009perception} but sometimes difficult \cite{richardson2010acoustic}. Scott Storch, a renowned music producer, has emphasized the importance of fine-tuning drums to match the music's key despite the inherent challenges in doing so.

In Section~\ref{subsec:808andproduction}, we showed that the Roland TR-808 Rhythm Composer has been deemed an influential analog drum machine \cite{meyers2003tr,werner2014physically,hasnain2017tr}, primarily known for its distinctive and deep bass drum sound \cite{carter1997tr}. Producers and musicians from various music genres have testified to its efficiency in providing low-end foundation. They use the 808 bass drum not only as a kick drum but also as a tonal instrument that plays basslines, thus emphasizing the importance of its tuning. 

\newpage

Signal analyses of 808 bass drum samples reported in Section~\ref{sec:808bassdrum} show that its fundamental frequency can be found ca. 50Hz and may or may not have lasting harmonics. The measured evolution of the power spectrum in popular music suggests that digital audio technology enabled the faithful reproduction of the 808 bass drum's extended bandwidth, which played a crucial role in the rise of trap music's popularity and its subsequent influence on mainstream music \cite{kaluvza2018reality}. 

In Section~\ref{sec:tuningto}, we discussed tuning the song's key to the 808 bass drum. Producers often try to tune the bass drum to match the song's key. However, Scott Storch suggests an alternative approach: adjusting the song's key to fit the 808 bass drum sample. Storch explains that some songs might have a key that is too low for the bass to be correctly reproduced by speakers. Instead, he recommends transposing the music up to achieve a more balanced and powerful bass response. 
If, for instance, the bass drum is transposed down one perfect fourth to match the song's key, its fundamental frequency loses ca. 11.8 dB in overall gain, considering the response of near field loudspeakers of the type that producers customarily use \cite{newell2001yamaha,senior2011mixing} and the human ear's sensitivity to frequency \cite{iso2262003}. The loss may affect the instrument's timbre and invalidate the specific bodily sensations the sub-bass range may evoke \cite{fink2018relentless,fink2020boom,hove2020feel}. The analysis also suggests that the gain loss primarily affects the fundamental frequency and lower harmonics. 
The discussion emphasizes the importance of controlling the level of bass in music production. It suggests that adjusting the song's key to the 808 bass drum can indeed be a helpful technique to achieve this goal.

In Section~\ref{sec:pitchandregister}, we briefly discussed the relationship between pitch and register in music and how transposition may affect these elements. While classical Western music theory emphasizes the robustness of pitch intervals to transposition, other perspectives \cite{frisius2010search,reymore2023timbre} suggest that transposition has significant consequences on timbre. Orchestration treatises have long associated specific timbral characteristics with different keys, highlighting the importance of considering both pitch and register \cite{berlioz1844grand,gevaert1885nouveau,koechlin1954traite}. Recent interviews with popular music producers suggest the approach is significant in modern music production.

An intriguing research direction may stem from the assessment of one of the interviewees, according to which spectral formants have precedence over pitch values in modern popular music. Storch's handling of the 808 bass drum is an example of this principle. If such a claim proves to have merit, it may have consequences on music analysis and user interaction in generative systems applied to popular music.







\section{Acknowledgments}


Many thanks to Yann Mac\'e and Luc Leroy from the music production company Hyper Music for their insights into Scott Storch's work and the subsequent discussions. Special thanks to David Meredith (Aalborg University) for his valuable comments.

\nocite{*}
\bibliography{ISMIRtemplate}

@article{antunes2017possible,
  title={Is it possible to tune a drum?},
  author={Antunes, Pedro RS},
  journal={Journal of Computational Physics},
  volume={338},
  pages={91--106},
  year={2017},
  publisher={Elsevier}
}

@MISC{Attack2022,
  author = {{Attack Magazine}},
  title  = {Four Of The Best Stem Separation Tools},
  howpublished = {https://youtu.be/9oNHoE4wHc8\&t=870 and t=1181},
  year = {2022},
  note = {accessed: 2023-11-09}
}

@MISC{BEA,
  author = {{Best Ever Albums}},
  title  = {Best ever albums},
  howpublished = {https://www.besteveralbums.com/},
  year = {2023},
  note = {accessed: 2023-11-09}
}

@book{berlioz1844grand,
  title={Grand trait{\'e} d'instrumentation et d'orchestration modernes, d{\'e}di{\'e} {\`a} sa majest{\'e} Fr{\'e}d{\'e}ric Guillaume IV roi de Prusse},
  author={Berlioz, Hector},
  year={1844},
  publisher={Schonenberger}
}

@MISC{burchell2022,
  author = {Burchell, Charles},
  title  = {Production Hacks: Creating 808 Basslines},
  howpublished = {https://articles.roland.com/production-hacks-creating-808-basslines/},
  year = {2022},
  month = {Nov.},
  note = {accessed: 2023-11-09}
}

@article{carter1997tr,
  title={{Roland TR808 Rhythm Composer (Retro)}},
  author={Carter, Chris},
  journal={Sound on Sound},
  year={1997},
  month ={May}
}

@article{chapman2008ill,
  title={{`That ill, tight sound': telepresence and biopolitics in post-Timbaland rap production}},
  author={Chapman, Dale},
  journal={Journal of the Society for American Music},
  volume={2},
  number={2},
  pages={155--175},
  year={2008},
  publisher={Cambridge University Press}
}

@article{dayal2014tr,
  title={{Roland TR-808}},
  author={Dayal, Geeta},
  journal={Grove Music Online},
  publisher = {Oxford Music Online},
  year={2014},
  month ={Jan.},
  url={https://doi.org/10.1093/gmo/9781561592630.article.A2257229}
}

@MISC{drummagazine,
  author = {{Drum Magazine}},
  title  = {How To Tune Drums In Four Steps},
  howpublished = {https://drummagazine.com/how-to-tune-drums-in-four-steps/},
  note = {accessed: 2023-11-09}
}

@MISC{Dunn2015,
  author = {Dunn, Alexander},
  title  = {808},
  howpublished = {https://youtu.be/KClqn0oN1lY},
  note = {accessed: 2023-11-09},
  year = {2015}
}

@book{fink2018relentless,
  title={The relentless pursuit of tone: Timbre in popular music},
  author={Fink, Robert and Latour, Melinda and Wallmark, Zachary},
  year={2018},
  publisher={Oxford University Press}
}

@article{fink2020boom,
  title={The Boom in the Box: Bass and Sub-Bass in Desktop Production},
  author={Fink, Robert},
  journal={The Bloomsbury Handbook of Music Production},
  year={2020},
  publisher={Bloomsbury Publishing USA}
}

@inproceedings{fitzgerald2010harmonic,
  title={Harmonic/percussive separation using median filtering},
  author={Fitzgerald, Derry},
  booktitle={13th International Conference on Digital Audio Effects (DAFX10), Graz, Austria},
  year={2010}
}

@article{fletcher1933loudness,
  title={Loudness, its definition, measurement and calculation},
  author={Fletcher, Harvey and Munson, Wilden A},
  journal={Bell System Technical Journal},
  volume={12},
  number={4},
  pages={377--430},
  year={1933},
  publisher={Wiley Online Library}
}

@article{frisius2010search,
  title={In Search of Lost Harmony},
  author={Frisius, Rudolf},
  journal={Contemporary music: theoretical and philosophical perspectives},
  pages={77--87},
  year={2010}
}

@book{gevaert1885nouveau,
  title={Nouveau trait{\'e} d'instrumentation},
  author={Gevaert, Fran{\c{c}}ois Auguste},
  year={1885},
  publisher={Lemoine \& fils}
}

@article{hasnain2017tr,
  title={How the {Roland TR-808} revolutionized music},
  author={Hasnain, Zainab},
  journal={The Verge},
  year={2017},
  month ={Apr.}
}

@article{hove2019increased,
  title={Increased levels of bass in popular music recordings 1955--2016 and their relation to loudness},
  author={Hove, Michael J and Vuust, Peter and Stupacher, Jan},
  journal={The Journal of the Acoustical Society of America},
  volume={145},
  number={4},
  pages={2247--2253},
  year={2019},
  publisher={AIP Publishing}
}

@article{hove2020feel,
  title={Feel the bass: Music presented to tactile and auditory modalities increases aesthetic appreciation and body movement.},
  author={Hove, Michael J and Martinez, Steven A and Stupacher, Jan},
  journal={Journal of Experimental Psychology: General},
  volume={149},
  number={6},
  pages={1137},
  year={2020},
  publisher={American Psychological Association}
}

@article{iso2262003,
  title={{Normal equal-loudness level contours-ISO 226: 2003}},
  author={ISO},
  year={2003},
  url={https://www.iso.org/standard/34222.html}
}

@article{kaluvza2018reality,
  title={Reality of Trap: Trap Music and its Emancipatory Potential.},
  author={Kalu{\v{z}}a, Jernej},
  journal={IAFOR Journal of Media, Communication \& Film},
  volume={5},
  number={1},
  year={2018}
}

@book{koechlin1954traite,
  title={Trait{\'e} de l'orchestration en 4 volumes},
  author={Koechlin, Charles},
  year={1954},
  publisher={Eschig}
}

@MISC{lavoie2020,
  author = {Lavoie, Alex},
  title  = {What is an 808? 7 Ways to Make Huge 808 Kicks},
  howpublished = {https://blog.landr.com/what-is-an-808/},
  note = {accessed: 2023-11-09},
  year = {2020},
  month = {Sep.}
}

@MISC{Lee2017,
  author = {Lee, Christina},
  title  = {{2 Chainz} Explains Why `Pretty Girls Like Trap Music'},
  howpublished = {https://www.rollingstone.com/music/music-features/2-chainz-explains-why-pretty-girls-like-trap-music-talks-his-bucket-list-and-benihana-193850},
  year={2017},
  note = {Accessed: 2023-11-09}
}

@MISC{Levine2003,
  author = {Levine, Robert and Werde, Bill},
  title  = {Superproducers},
  howpublished = {https://www.wired.com/2003/10/producers/},
  note = {accessed: 2023-11-09},
  year = {2003},
  month = {Oct.}
}

@article{licklider1951duplex,
  title={A duplex theory of pitch perception},
  author={Licklider, Joseph Carl Robnett},
  journal={The Journal of the Acoustical Society of America},
  volume={23},
  number={1, Supplement},
  pages={147--147},
  year={1951},
  publisher={AIP Publishing}
}

@MISC{Loose2022,
  author = {Loose, Sam},
  title  = {{iZotope RX} 9 - Fixing Drums In The Master??},
  howpublished = {https://youtu.be/LCH23ZiTXCA\&t=177},
  year = {2022},
  note = {accessed: 2023-11-09}
}

@techreport{meyers2003tr,
  title={{Roland TR-808 Rhythm Composer}},
  author={Meyers, Owen},
  institution={McGill University},
  year={2003}
}

@book{millard2005america,
  title={America on record: a history of recorded sound},
  author={Millard, Andre},
  year={2005},
  publisher={Cambridge University Press}
}

@article{newell2001yamaha,
  title={The {Yamaha NS10M}: twenty years a reference monitor. {W}hy?},
  author={Newell, Philip R and Holland, Keith R and Newell, Julius P},
  journal={Proceedings of the Institute of Acoustics},
  volume={23},
  number={8},
  pages={29--40},
  year={2001}
}

@inproceedings{ono2008real,
  title={A Real-time Equalizer of Harmonic and Percussive Components in Music Signals.},
  author={Ono, Nobutaka and Miyamoto, Kenichi and Kameoka, Hirokazu and Sagayama, Shigeki},
  booktitle={ISMIR},
  pages={139--144},
  year={2008}
}

@inproceedings{pestana2013spectral,
  title={Spectral characteristics of popular commercial recordings 1950-2010},
  author={Pestana, Pedro Duarte and Ma, Zheng and Reiss, Joshua D and Barbosa, Alvaro and Black, Dawn AA},
  booktitle={Audio Engineering Society Convention 135},
  year={2013},
  organization={Audio Engineering Society}
}

@book{read1952reproduction,
    title={The Recording and Reproduction of Sound. A Complete Reference Manual for the Professional and the Amateur, 2nd edition},
    author={Read, Oliver},
     year={1952},
    publisher={Howard W. Sams \& Co.}
}

@article{reid2002bass,
  title={Practical Bass Drum Synthesis},
  author={Reid, Gordon},
  journal={Sound on Sound},
  year={2002},
  month={Feb.}
}

@article{reymore2023timbre,
  title={Timbre Semantic Associations Vary Both Between and Within Instruments: An Empirical Study Incorporating Register and Pitch Height},
  author={Reymore, Lindsey and Noble, Jason and Saitis, Charalampos and Traube, Caroline and Wallmark, Zachary},
  journal={Music Perception: An Interdisciplinary Journal},
  volume={40},
  number={3},
  pages={253--274},
  year={2023},
  publisher={University of California Press}
}

@phdthesis{richardson2010acoustic,
  title={Acoustic analysis and tuning of cylindrical membranophones},
  author={Richardson, Philip GM},
  year={2010},
  school={Anglia Ruskin University}
}

@MISC{Roberts,
  author = {Roberts, John},
  title  = {About drums},
  howpublished = {https://circularscience.com /about-drums/},
  note = {accessed: 2023-11-09}
}

@article{Robinson1997,
  author = {Robinson, Andrea},
  title  = {RADIO DAYS},
  journal = {The Mix},
  year = {1997},
  month = {Aug.}
}

@inproceedings{rump2010autoregressive,
  title={Autoregressive {MFCC} Models for Genre Classification Improved by Harmonic-percussion Separation.},
  author={Rump, Halfdan and Miyabe, Shigeki and Tsunoo, Emiru and Ono, Nobutaka and Sagayama, Shigeki},
  booktitle={ISMIR},
  pages={87--92},
  year={2010}
}

@MISC{Stemroller2023,
  author = {{Verysickbeats}},
  title  = {{This Might be the CLEANEST Stem Remover from Songs | StemRoller}},
  howpublished = {https://youtu.be/-G76oQ3uL90\&t=364},
  year = {2023},
  note = {accessed: 2023-11-09}
}

@article{takahashi2002relationship,
  title={The relationship between vibratory sensation and body surface vibration induced by low-frequency noise},
  author={Takahashi, Yukio and Kanada, Kazuo and Yonekawa, Yoshiharu},
  journal={Journal of low frequency noise, vibration and active control},
  volume={21},
  number={2},
  pages={87--100},
  year={2002},
  publisher={SAGE Publications Sage UK: London, England}
}

@book{senior2011mixing,
  title={Mixing secrets for the small studio},
  author={Senior, Mike},
  year={2011},
  publisher={Taylor \& Francis}
}

@MISC{Singleton2022,
  author = {Singleton, Mya},
  title  = {{20 producers that  the R\&B game}},
  howpublished = {https://www.yardbarker.com/entertainment/articles/
  20\_producers\_that\_\_the\_r\_b\_game/s1\_\_37845173},
  note = {accessed: 2023-11-09},
  year = {2022},
  month = {Oct.}
}

@MISC{Storch2022,
  author = {Storch, Scott},
  title  = {Masterclass: becoming a hitmaker with {Scott Storch}. {C}hapter 8, `{D}efining a bass line'},
  howpublished = {https://www.aulart.com/masterclass/scott-storch-becoming-a-hitmaker/},
  note = {accessed: 2023-11-09},
  year = {2022}
}

@MISC{Storch2022b,
  author = {Storch, Scott},
  title  = {Masterclass: becoming a hitmaker with {Scott Storch}. {C}hapter 6, `{S}electing drum sounds'},
  howpublished = {https://www.aulart.com/masterclass/scott-storch-becoming-a-hitmaker/},
  note = {accessed: 2023-11-09},
  year = {2022}
}

@MISC{storch2007,
  author = {{Penton Media}},
  title  = {Scott Storch on Kick Drums},
  howpublished = {https://youtu.be/R5PUcMAFZOI},
  year = {2007},
  note = {accessed: 2023-11-09}
}

@article{toulson2009perception,
  title={The perception and importance of drum tuning in live performance and music production},
  author={Toulson, Rob and Crigny, Charles Cuny and Robinson, Philip and Richardson, Phillip},
  journal={The Journal on the Art of Record Production},
  volume={4},
  year={2009}
}

@MISC{trdownload,
  author = {Trisamples},
  title  = {{TR-808}},
  howpublished = {https://trisamples.com/roland-tr808-free-download/},
  year = {2020},
  note = {accessed: 2023-11-09}
}

@inproceedings{werner2014physically,
  title={A Physically-Informed, Circuit-Bendable, Digital Model of the {Roland TR-808} Bass Drum Circuit.},
  author={Werner, Kurt James and Abel, Jonathan S and Smith III, Julius O},
  booktitle={DAFx},
  pages={159--166},
  year={2014}
}

@article{wu2018review,
  title={A review of automatic drum transcription},
  author={Wu, Chih-Wei and Dittmar, Christian and Southall, Carl and Vogl, Richard and Widmer, Gerhard and Hockman, Jason and M{\"u}ller, Meinard and Lerch, Alexander},
  journal={IEEE/ACM Transactions on Audio, Speech, and Language Processing},
  volume={26},
  number={9},
  pages={1457--1483},
  year={2018},
  publisher={IEEE}
}

@inproceedings{yoo2010nonnegative,
  title={Nonnegative matrix partial co-factorization for drum source separation},
  author={Yoo, Jiho and Kim, Minje and Kang, Kyeongok and Choi, Seungjin},
  booktitle={2010 IEEE International Conference on Acoustics, Speech and Signal Processing},
  pages={1942--1945},
  year={2010},
  organization={IEEE}
}

@MISC{zisook2007,
  author = {Zisook, Brian},
  year={2007},
  title  = {{Remix Hotel Atlanta} Announces Keynote Speaker {Scott Storch}},
  howpublished = {https://djbooth.net/features/remix-hotel-atlanta-announces-keynote-speaker-scott-storch-0830072},
  note = {accessed: 2023-11-09}
}

%
%
%
%
%

\end{document}